\documentclass[aps,prl,twocolumn,showpacs]{revtex4}
\usepackage{bm}
\usepackage{epsfig}
\usepackage{graphicx}
\usepackage{amssymb,amsmath,amsbsy,amsgen,amsfonts}     
\usepackage{dcolumn}
\usepackage{amsthm}
\usepackage{mathrsfs}
\usepackage{latexsym}
\usepackage{array}
\usepackage{amstext}

\newcommand{\ncd}{\newcommand}
\ncd{\QC}{$\mbox{QC}_{\cal{C}}$}
\ncd{\QCpr}{${\mbox{QC}_{\cal{C}}}^\prime\;$}
\ncd{\QCns}{$\mbox{QC}_{\cal{C}}$}
\ncd{\QCprns}{${\mbox{QC}_{\cal{C}}}^\prime$}
\ncd{\cskN}{{|\phi_{\{\kappa\} } \rangle}_{{\cal{C}}_N}}
\ncd{\cskNpr}{{|\phi_{\{\kappa^\prime\} } \rangle}_{{\cal{C}}_N}}
\ncd{\cskNtil}{{|\phi_{\{\tilde{\kappa} \} }
\rangle}_{{\cal{C}}_N}}
\ncd{\csk}{{|\phi_{\{\kappa\} }
\rangle}_{\cal{C}}}
\ncd{\csktil}{{|\phi_{\{\tilde{\kappa} \} }
\rangle}_{\cal{C}}}
\ncd{\cskf}{|\phi_{\{\kappa\} }
\rangle_{\cal{C}}}
\ncd{\csktilf}{|\phi_{\{\tilde{\kappa} \} }
\rangle_{\cal{C}}}
\ncd{\bracsk}{\mbox{}_{\cal{C}}\langle\phi_{\{\kappa\} }|}
\ncd{\bracsktil}{\mbox{}_{\cal{C}}\langle\phi_{\{\tilde{\kappa} \}
}|} \ncd{\nbracsk}{\mbox{}_{\cal{C}}\langle\phi_{\{\kappa\} }}
\ncd{\nbracsktil}{\mbox{}_{\cal{C}}\langle\phi_{\{\tilde{\kappa}
\} }} \ncd{\cs}{|\phi \rangle_{\cal{C}}\;} \ncd{\csns}{|\phi
\rangle_{\cal{C}}}
\ncd{\nbgh}{\text{nghb}} 
\ncd{\Sab}{S^{ab}}
\ncd{\Sba}{S^{ba}}
\ncd{\ds}{\displaystyle} \ncd{\ovl}{\overline}

\ncd{\SABC}{S^{ABC}}
\ncd{\Sbc}{S^{bc}}

\newcommand{\ket}[1]{\left\vert{#1}\right\rangle}
\newcommand{\ketbra}[2]{|#1\rangle \langle#2|}

\newcommand{\be}{\begin{equation}}
\newcommand{\ee}{\end{equation}}
\newcommand{\ba}{\begin{array}}
\newcommand{\ea}{\end{array}}
\newcommand{\bqa}{\begin{eqnarray}}
\newcommand{\eqa}{\end{eqnarray}}

\begin{document}

\title{Compact Toffoli gate using weighted graph states}
\author{M. S. Tame,$^{1,2}$ \c{S}. K. \"Ozdemir,$^{2,3,4}$ M. Koashi,$^{2,4}$ N. Imoto,$^{2,4}$ and M. S. Kim$^1$} 
\affiliation{$^1$School of Mathematics and Physics, The Queen's University, Belfast, BT7 1NN, UK\\
$^2$Graduate School of Engineering Science, Osaka University, Toyonaka, Osaka 560-8531, Japan\\
$^3$Department\,of\,Electrical\,and\,Systems\,Engineering,\,Washington University,\,St.\,Louis,\,Missouri\,63130,\,USA\\
$^4$CREST\,Research\,Team\,for\,Photonic\,Quantum\,Information,\,Saitama\,331-0012,\,Japan}

\date{\today}

\begin{abstract}
We introduce three compact graph states that can be used to perform a measurement-based Toffoli gate. Given a weighted graph of six, seven or eight qubits, we show that success probabilities of $1/4$, $1/2$ and $1$ respectively can be achieved. Our study puts a measurement-based version of this important quantum logic gate within the reach of current experiments. As the graphs are setup-independent, they could be realized in a variety of systems, including linear optics and ion-traps.    
\end{abstract}


\maketitle

While there has been steady progress in experimentally demonstrating single and two-qubit logic gates for quantum computation~(QC)~\cite{nielsenchuang}, combining these elements for performing useful algorithms is still far too demanding. Grover's and Shor's algorithms~\cite{nielsenchuang,GrovShor}, for example, require highly {\it non-local} logic gates during their operation~\cite{nielsenchuang}. In this context, an important building block is the three-qubit {\it Toffoli} gate~\cite{ToffoliC5,Barenco}; it is universal, {\it i.e.} together with single-qubit gates one can perform any QC, with the advantage that one can efficiently build highly non-local $n$-time control-{\sf NOT} gates with a polynomial dependence on $n$ for the number of Toffoli gates required~\cite{ToffoliC5}.
Recently, a reduction in the resources required to implement this gate has been proposed~\cite{FiuRalph} and experimentally realized~\cite{MonzLanyon}. However, this work is limited to the standard network model for QC~\cite{nielsenchuang}.
A promising alternative to the network model is the {\it measurement-based} one-way model~\cite{oneway}. This has attracted much interest recently, due to its advantages over the network model in reducing the level of control required and increasing flexibility for a variety of physical systems~\cite{oneway,onewayt,onewaye}. 
Here, we introduce a 
measurement-based Toffoli gate that can be implemented with current technology.
We address an important issue in QC, providing a practical alternative to existing schemes for a range of physical systems. 

In this work we introduce three compact graph states for implementing a Toffoli gate using one-way QC~\cite{oneway}; computations are performed by making single-qubit measurements on an initial entangled resource. Usually the resource used is a form of cluster state~\cite{oneway}, however, more general graph states~\cite{graph} can also be employed~\cite{BrowneBriegel}. Here, in contrast to earlier work, we show that by incorporating {\it weighted} edges in graph states of six, seven or eight qubits, one can construct compact Toffoli gates with success probabilities of 1/4, 1/2 and 1 respectively. The smallest graph previously known to achieve a Toffoli gate consisted of at least ten qubits and a complex entanglement structure~\cite{BrowneBriegel}, making it too challenging for current experiments~\cite{onewaye}. Our study puts a measurement-based version of this important gate within reach of current technology. To emphasize this, we provide a basic example for generating the six-qubit graph using linear optics. However, as the graphs are setup-independent, we expect them to be useful in many other physical systems. Before providing details of how our proposed Toffoli gate works, we give a brief review of one-way QC using weighted graphs and the basic tools needed for its understanding.

{\it One-way QC with weights.-} A weighted graph state $\ket{G}$ is a multipartite entangled state consisting of a set of vertices $j$ (qubits prepared in $\ket{+}_j$, where $\ket{\pm}_j= \frac{1}{\sqrt{2}}(\ket{0} \pm \ket{1})_j$ and $\{{\left| 0 \right\rangle}_{j},{{\left|1\right\rangle} }_{j}\}$ is the single-qubit computational basis) connected to each other by edges taking the form of entangling operations ${\sf CZ}^\theta=diag(1,1,1,e^{i \theta_{jk}})$~\cite{graph}. Here, the {\it weights} $\theta_{jk} \in [0,2\pi]$ correspond to a controlled phase operation applied between qubits $j$ and $k$. For standard one-way QC on maximally weighted graphs ($\theta_{jk}=\pi, \forall~(i,j) \in G$), two essential types of single-qubit measurements are involved~\cite{oneway, eisert}. First, measuring the state of qubit $j$ in the computational basis disentangles it from the graph, leaving a smaller entangled resource. This allows one to shape a given graph for a particular processing task. Second, in order to perform QC, qubits can be measured in the basis $B_{j}(\alpha)=\{\mathop{{\left|\alpha_{\pm}\right\rangle}_{j}}=\frac{1}{\sqrt 2}({\left|0\right\rangle}\pm{e}^{i\alpha}{\left|1\right\rangle})_{j}\}$ ($\alpha\in\mathbb{R}$). This applies a single-qubit rotation $R_{z}^{-\alpha}=diag(1,{\rm e}^{-i\alpha})$, followed by a Hadamard operation ${\sf H}=(\sigma_x + \sigma_z)/\sqrt{2}$ to a {\it logical} qubit residing on vertex $j$ ($\sigma_{x,y,z}$ are the Pauli matrices). At the same time, this logical qubit is transferred to the next {\it available} adjacent vertex, {\it i.e.} a neighboring vertex without a logical qubit already residing on it~\cite{BrowneBriegel}. With proper choices for the $B_{j}(\alpha)$'s and an appropriate configuration for the graph~\cite{oneway,Nest},
any quantum circuit can be performed. However, once the weights of edges become non-maximal, care must be taken in the order that measurements are made~\cite{eisert}. In this work we consider weighted edges {\it between} logical qubits residing on qubits in the graph and not {\it along} their path of flow.
\begin{figure}[b]
\centerline{\psfig{figure=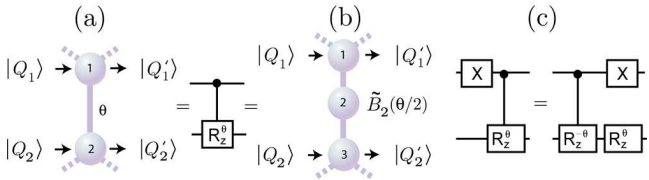,width=8.5cm}}
\caption{{\sf CZ}$^\theta$ gates. Edges are maximal unless marked. {\bf (a)}: No measurements are made to perform the {\sf CZ}$^\theta$. {\bf (b)}: $\tilde{B}_2(\theta/2)$ is used to implement {\sf CZ}$^\theta$ up to local terms~\cite{BrowneBriegel}. {\bf (c)}: Passage of $\sigma_x$ ($={\sf X}$) through the gate carried out by the resource in panel {\bf (a)}. The gate acts symmetrically on $\sigma_x$ byproducts.}
\label{identities}
\end{figure}

In Fig.~\ref{identities}~{\bf (a)} we show a simple two-qubit weighted graph state, with two {\it logical} qubits $\ket{Q_{1}}$ and $\ket{Q_{2}}$ residing on qubits 1 and 2 respectively. Here the edge connecting the qubits has the weight $\theta$. No measurements are required in this case and the resource simulates a ${\sf CZ}^\theta$ gate. This graph can be {\it linked} with a larger graph by connecting up each qubit with maximally weighted edges: one edge {\it in} and one edge {\it out}, enabling qubit information flow (shown as dashed lines in Fig.~\ref{identities}). Due to the probabilistic nature of the outcomes from measurements that are used to transfer logical qubits across a given resource, byproduct Pauli operators are generated. 
For standard one-way QC, these must pass {\it freely} through the quantum circuit generated by subsequent measurements, {\it i.e.} they produce no change to the overall computation when passed through to the end of the circuit, where they can be removed~\cite{oneway}. This ensures that the entire computation remains unchanged and therefore deterministic. Unfortunately, in Fig.~\ref{identities}~{\bf (a)} only $\sigma_z$ byproducts will pass through this part of the circuit freely. In Fig.~\ref{identities}~{\bf (c)} we show the effect of passing $\sigma_x$ through ${\sf CZ}^\theta$. The flip operation, {\it i.e.} ${\sf CZ}^{\theta} \to {\sf CZ}^{-\theta}$, and the additional non-Pauli byproduct $R_{z}^{\theta}$ can cause problems and reduce the overall success probability if not accommodated for properly. We will show how they can be integrated into a compact measurement-based Toffoli gate in the next section. In Fig.~\ref{identities}~{\bf (b)}, we show a three-qubit graph, with $\ket{Q_{1}}$ and $\ket{Q_{2}}$ residing on qubits 1 and 3 respectively. This graph is also important for the Toffoli gate operation. Here $\tilde{B}_2(\theta/2):={\sf H} B_2(\theta/2)$ is used and the resource performs the gate $[R_z^{-\theta/2}\otimes R_z^{-\theta/2}]{\sf CZ}^{\theta}$~\cite{BrowneBriegel}, with byproduct $\Sigma_{\cal M}=\sigma_z^{s_2} \otimes \sigma_z^{s_2}$ ($s_i \in \{0,1 \}$ corresponds to the outcome of the measurement on qubit $i$). This graph can also be linked to a larger graph with the central qubit providing the freedom to choose $\theta$ after the resource is generated. In addition, one can pass $\sigma_x$ through without flipping the sign of the desired gate as follows: The most general form of  {\it linking} byproducts, {\it i.e.} those generated from previous parts of a computation in the larger graph, is $\Sigma_{\cal L}=[\sigma_x^{s^x_{Q_1}}\sigma_z^{s^z_{Q_1}} \otimes \sigma_x^{s^x_{Q_2}}\sigma_z^{s^z_{Q_2}}]_{Q_1Q_2}$. The values $s_{Q_i}^{x,z}\in \{0,1\}$ correspond to the accumulation of $\sigma_{x,z}$ byproducts on logical qubit $\ket{Q_i}$. $\Sigma_{\cal L}$ can be passed through by {\it absorbing} the operation ${\cal R}=(\sigma_z^{s^x_{Q_1} \oplus s^x_{Q_2}})$ into $\tilde{B}_2$, {\it i.e.} the measurement basis is adapted: $\tilde{B}_2 \to {\cal R}^{\dag}\tilde{B}_2$.
\begin{figure}[t]
\centerline{\psfig{figure=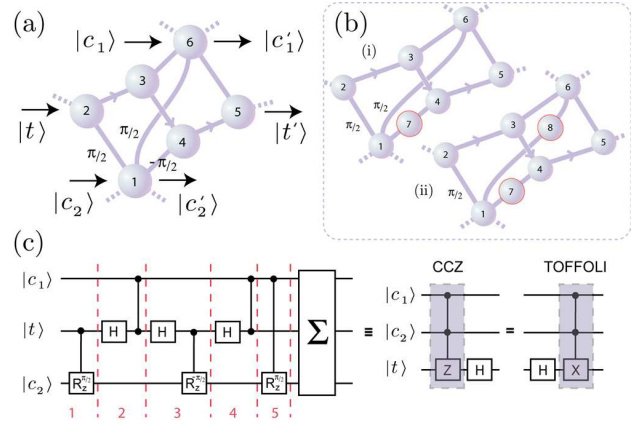,width=8.2cm}}
\caption{Compact graph states for a Toffoli gate. {\bf (a)}: Six-qubit graph. {\bf (b)}: (i) Seven and (ii) Eight-qubit graphs. {\bf (c)}: Network circuit induced by measurements of qubits $\{2,3,4\}$, $\{2,3,4,7\}$ and $\{2,3,4,7,8\}$ for the six, seven, and eight-qubit graphs respectively. $\Sigma$ denotes a combination of byproducts from {\it measurement} outcomes and previous byproducts from {\it linking} into larger graphs (dashed lines in {\bf (a)} and {\bf (b)}). The right hand side shows the equivalence to a Toffoli gate, where the additional {\sf H} can be removed as described in the text.}
\label{config}
\end{figure}    

{\it Compact Toffoli gate.-} With the above considerations, we now show how six, seven and eight-qubit graph states can perform a compact Toffoli gate, together with an explanation of their success probabilities. The graphs were found by a direct mapping of the most compact quantum circuit known for the Toffoli gate~\cite{nielsenchuang,Marg}, in terms of controlled-{\sf NOT} and ${\sf CZ}^{\theta}$ gates, to the basic building block resources used in the one-way model~\cite{oneway,onewayt}. For simplicity, we will describe only the relation between the final graphs constructed and their corresponding circuits. We start with the operation of the six-qubit graph when there are no byproducts. In Fig.~\ref{config}~{\bf (a)} the in/out logical control qubits $\ket{c_1}$ and $\ket{c_2}$ reside on qubits 6 and 1 respectively and the input target qubit $\ket{t}$ resides on qubit 2. A ${\sf CZ}^{\pi/2}$ gate is initially applied between $\ket{t}$ and $\ket{c_2}$, corresponding to the edge of weight $\pi/2$ linking qubits 1 and 2. This operation is depicted in Fig.~\ref{config}~{\bf (c)} as {\it step 1}. Next, qubit 2 is measured in the basis $B_2(0)$. This applies {\sf H} to $\ket{t}$ while propagating it across to reside on qubit 3, where a {\sf CZ} gate is applied between $\ket{c_1}$ and $\ket{t}$. These operations are shown as {\it step 2}. This is followed by a measurement of qubit 3 in the basis $B_3(0)$, which applies {\sf H} to $\ket{t}$ and propagates it across to reside on qubit 4, where a ${\sf CZ}^{-\pi/2}$ gate is applied between $\ket{t}$ and $\ket{c_2}$. These operations are depicted as {\it step 3}. Next, qubit 4 is measured in the basis $B_4(0)$. This applies {\sf H} to $\ket{t}$, while propagating it to reside on qubit 5, where a {\sf CZ} gate is applied between $\ket{c_1}$ and $\ket{t}$. These operations are shown as {\it step 4}. Finally, the weighted edge between qubits 1 and 6 is included, which performs a ${\sf CZ}^{\pi/2}$ gate between $\ket{c_1}$ and $\ket{c_2}$. This is given in Fig.~\ref{config}~{\bf (c)} as {\it step 5}. The overall circuit is equivalent to a Toffoli gate~\cite{nielsenchuang,Marg}, up to a local {\sf H} on the input target. This can be removed by encoding $\ket{t}$ into the graph in the {\sf H} basis. Alternatively, reversing the direction of qubit flow puts the {\sf H} on the output target. This could be removed by {\it feed-forward} operations, such as those used in linear optics~\cite{onewaye}. We note that if the weights $\pm \pi/2 \to \pm \theta/2$, the graph can be used as a control-control-{\sf Z} phase operation (${\sf CCZ}^{\theta}$)~\cite{nielsenchuang,Marg}.

{\it Success probability.-} We now discuss the important role of byproducts in above procedure. The operator $\Sigma$ shown in Fig.~\ref{config}~{\bf (c)} is a combination of the byproduct $\Sigma_{{\cal M}_6}$ resulting from measurements of qubits 2, 3 and 4, with a passed-through $\Sigma_{\cal L}$ 
resulting from previous measurements in a larger graph linked up to the six-qubit graph. First, it is straightforward to show $\Sigma_{{\cal M}_6}=[(\sigma_z^{s_4})_{c_1}\otimes (R_z^{-\pi/2})^{s_3}_{c_2}\otimes (\sigma_x^{s_2 \oplus s_4}\sigma_z^{s_3})_{t}][{\sf CNOT}_{c_2t}{\sf CZ}_{c_1c_2}]^{s_3}$. Thus, half of these byproducts ($s_3=0$) consist of local Pauli operations and the other half ($s_3=1$) consist of non-local operations. This results in a success probability of $p_s=1/2$, as $\Sigma_{{\cal M}_6}$ for $s_3=0$ can be passed through until the end of a computation. However, $\Sigma_{\cal L}$ also needs to be taken into account. The most general form of $\Sigma_{\cal L}$ for the logical {\it input} qubits is $[\sigma_x^{s^x_{c_1}}\sigma_z^{s^z_{c_1}}\otimes\sigma_x^{s^x_{c_2}}\sigma_z^{s^z_{c_2}}\otimes\sigma_x^{s^x_{t}}\sigma_z^{s^z_{t}}]_{c_1 c_2 t}$. As $\sigma_z$'s pass freely through the circuit of Fig.~\ref{config}~{\bf (c)} (with $\sigma_z \to \sigma_x$ on the target line), they do not lower $p_s$. 
On the other hand, $\sigma_x$'s modify the ${\sf CZ}^{\pm \pi/2}$ gates, as shown in Fig.~\ref{identities}~{\bf (c)}. Fortunately, for a selection of cases, $\sigma_x$'s can be passed through by adapting the measurement bases of the qubits in the graph, with $\Sigma_{{\cal L}}$ changing accordingly. Thus, we have $\Sigma=\tilde{\Sigma}_{\cal L}\Sigma_{{\cal M}_6}$, where $\sim$ denotes a possible change. 
By taking into account sign flips of weighted edges (see Fig.~\ref{identities}~{\bf (c)}), such that the phase of the second gate in Fig.~\ref{config}~{\bf (c)} is always opposite to the first and third, it is straightforward to show that for ${\cal S}^{x}:=\{s^x_{c_1},s^x_{c_2},s^x_{t}\}=\{ 0,0,0 \}$, $\{ 0,1,0 \}$, $\{ 1,0,1 \}$ and $\{ 1,1,1 \}$, one can pass through $\sigma_x$'s in $\Sigma_{\cal L}$ by absorbing into the measurement bases of the qubits in the graph the operations ${\cal R}_{000}=\openone$ ($\Sigma=\Sigma_{{\cal M}_6}$), ${\cal R}_{010}=(R_z^{-\pi/2})_2\otimes(R_z^{\pi/2})_4$ ($\Sigma=(R_z^{\pi/2}\sigma_z^{s_4})_{c_1}\otimes(\sigma_x)_{c_2}\otimes(\sigma_x^{s_2 \oplus s_4})_{t}$), ${\cal R}_{101}=(\sigma_z)_3$ ($\Sigma=(\sigma_x\sigma_z^{s_4})_{c_1}\otimes(R_z^{\pi/2})_{c_2}\otimes(\sigma_x^{s_2 \oplus s_4})_{t}$) and ${\cal R}_{111}=(\sigma_z)_3{\cal R}_{010}$ ($\Sigma=(\sigma_x R_z^{-\pi/2}\sigma_z^{s_4})_{c_1}\otimes(\sigma_xR_z^{-\pi/2})_{c_2}\otimes(\sigma_x^{s_2 \oplus s_4})_{t}$)~\cite{nonpauli}. Here $s_3=0$ is assumed. The remaining ${\cal S}^{x}$ cases lead to a non-local $\Sigma$. As each ${\cal S}^x$ occurs with probability 1/8~\cite{oneway,link} and $s_3=0$ with probability 1/2, the above four cases lead to a total success probability of $p_s=1/4$. If the six-qubit graph is linked with a larger graph, a careful design of the combined graph would enable all non-local $\Sigma$ to be removed (including the cases when $s_3=1$) by using {\it bridging} qubits between the paths of qubit flow~\cite{break}. This would give $p_s=1$. However, from a minimal-resource perspective, this requires a large additional overhead. We now show that with just one more qubit and an edge, one can obtain $p_s=1/2$. Then, with an extra two qubits and two edges one can achieve $p_s=1$. 
\begin{figure}[t]
\centerline{\psfig{figure=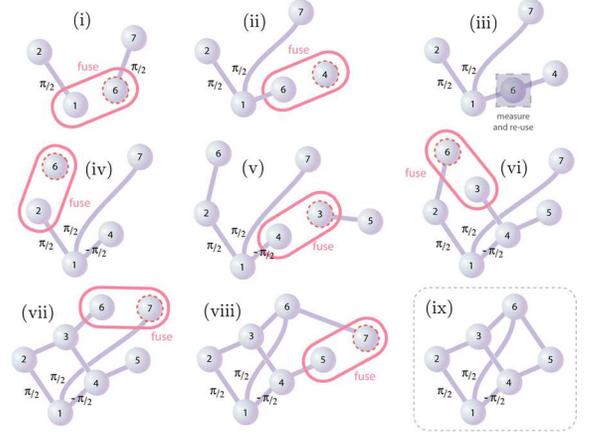,width=7.5cm}}
\caption{Generation of the six-qubit graph using linear optics. 
} 
\label{construct}
\end{figure}

{\it Seven and eight qubits.-} Consider the addition of an extra qubit between qubits 1 and 4, as shown in Fig.~\ref{config}~{\bf (b)}~(i), to make a seven-qubit graph. The {\it sub-graph} of qubits 1,~4 and 7 constitutes the three-qubit resource of Fig.~\ref{identities}~{\bf (b)}. Thus it can be used to implement a ${\sf CZ}^{-\pi/2}$ gate between $\ket{t}$ and $\ket{c_2}$. Here, $\tilde{B}_7(-\pi/4)$ is used, while $B_4(0) \to B_4(\pi/4)$. Furthermore, qubit 3 should be measured first, with ${\cal R}=[(\sigma_x)_4 \otimes (\sigma_z)_7]^{s_3}$ absorbed into the bases of qubits 4 and 7. The byproduct $\Sigma_{{\cal M}_7}=(\sigma_z^{s_4 \oplus s_7})_{c_1} \otimes (\sigma_z^{s_7}R_z^{\pi/4})_{c_2} \otimes (\sigma_z^{s_3}\sigma_x^{s_2\oplus s_4 \oplus s_7})_{t}$ is now completely local in form. By taking $\Sigma_{\cal L}$ into account also, we have ${\cal R}_{000}=\openone$ ($\Sigma=\Sigma_{{\cal M}_7}$), ${\cal R}_{010}=(R_z^{-\pi/2})_2 \otimes (\sigma_x)_4$ ($\Sigma=(R_z^{\pi/2})_{c_1} \otimes (\sigma_x R_z^{-\pi/2})_{c_2}\Sigma_{{\cal M}_7}$), ${\cal R}_{101}= (\sigma_x)_4 \otimes (\sigma_z)_7$ ($\Sigma=(\sigma_x)_{c_1} \otimes (R_z^{\pi/2})_{c_2}\otimes (\sigma_z)_{t}\Sigma_{{\cal M}_7}$) and ${\cal R}_{111}={\cal R}_{010}{\cal R}_{101}$ ($\Sigma=(\sigma_x R_z^{-\pi/2} )_{c_1} \otimes (\sigma_x \sigma_z)_{c_2}\otimes (\sigma_z)_{t}\Sigma_{{\cal M}_7}$). Thus, we have $p_s=1/2$. With an extra qubit between qubits 1 and 6, as shown in Fig.~\ref{config}~{\bf (b)}~(ii), to make an eight-qubit graph, all $\sigma_x$'s in $\Sigma_{\cal L}$ can be passed freely. Qubits 1, 6 and 8 apply a ${\sf CZ}^{\pi/2}$ gate on $\ket{c_1}$ and $\ket{c_2}$, with $\tilde{B}_8(\pi/4)$. We then have $\Sigma_{{\cal M}_8}=(\sigma_z^{s_4 \oplus s_7 \oplus s_8}R_z^{-\pi/4})_{c_1} \otimes (\sigma_z^{s_7 \oplus s_8})_{c_2} \otimes (\sigma_z^{s_3}\sigma_x^{s_2\oplus s_4 \oplus s_7})_{t}$, with $\sigma_x$'s from $\Sigma_{\cal L}$ passed freely by absorbing ${\cal R}_{ijk}=[(\sigma_z)_3 \otimes (\sigma_z)_8]^i[(R_z^{-\pi/2})_2 \otimes (\sigma_x)_4]^j[(\sigma_z)_8]^k$ and $\Sigma=[(\sigma_x)_{c_1}\otimes(\sigma_z)_t]^i[(R_z^{\pi/2})_{c_1}\otimes (\sigma_x)_{c_2}]^j [(R_z^{\pi/2})_{c_1}\otimes(R_z^{\pi/2})_{c_2}\otimes (\sigma_z)_t]^k [(\sigma_z)_{c_1}\otimes (\sigma_z)_{c_2}]^{jk}\Sigma_{{\cal M}_8}$. Here, contrary to the seven qubit case, ${\cal R}$ comes after the application of ${\cal R}_{ijk}$. We now have $p_s=1$.

{\it Linear optics demonstration.-}
As an example physical setup, we now describe how to generate the six-qubit graph in a linear optics setting.~The seven and eight-qubit resources can be constructed similarly. To create the six-qubit graph, the {\it steps} given in Fig.~\ref{construct} could be used. Here, polarization degrees of freedom of a photon in mode $i$ are used to embody a qubit, where $\ket{0}_i \to \ket{H}_i$ and $\ket{1}_i \to \ket{V}_i$. The {\it fuse} operation (depicted as a rounded square) is given by $\ketbra{HH}{HH}+\ketbra{VV}{VV}$, followed by {\sf H} on the qubit with a dotted ring. This is realized when a single photon exits each output port of a polarizing beamsplitter (PBS), on the condition that the two input photons entered seperate ports~\cite{onewayt,onewaye}. Single-qubit rotations such as {\sf H} and $R_z^{\alpha}$ can be implemented with half-wave plates (HWPs) and quarter-wave plates (QWPs) on the relevant photon modes. In {\it step}~(i), qubits 1 and 6 from two weighted graphs $(\ket{H}\ket{+}+\ket{V}\ket{\pi/2_+})_{21}\otimes (\ket{H}\ket{+}+\ket{V}\ket{\pi/2_+})_{67}$ are fused. These could each be produced from a concatenated parametric down-conversion type-I process~\cite{PDC}. With the laser pump polarization set correctly, the state $[\gamma_+\ket{HH}+\gamma_-\ket{VV}]_{ab}$ can be generated in output modes $a$ and $b$, where $\gamma_{\pm}=\frac{1}{2}(1\pm e^{-i \gamma/2})$. The operation $[R_z^{\gamma/2}{\sf H} \otimes R_z^{\gamma/2} {\sf H}]_{ab}$ rotates the state into a two-qubit graph with weighted edge $\gamma$. A double-pass scheme would generate both graphs from the same crystal. In {\it step}~(i), the result of the fusion is the state $[\ket{+}\ket{H}\ket{+}\ket{+}+\ket{\pi/2_+}\ket{V}\ket{-}\ket{\pi/2_+}]_{2167}$. Next, in {\it step}~(ii) a fuse operation is applied to modes 6 and 4 (initially set to $\ket{+}$). This produces the state $[\ket{+}\ket{H}(\ket{H}\ket{+}+\ket{V}\ket{-})\ket{+}+\ket{\pi/2_+}\ket{V}(\ket{H}\ket{+}-\ket{V}\ket{-})\ket{\pi/2_+}]_{21647}$. In {\it step}~(iii) $\tilde{B}_6(\pi/4)$ is used, this is achieved with a QWP, HWP and polarizer in the corresponding photon mode. The rotations $[R_z^{\pi/4}\otimes R_z^{\pi/4}]_{14}$ are then applied. The resulting state is given by $[\ket{+}\ket{H}\ket{+}\ket{+}+\ket{\pi/2_+}\ket{V}\ket{-\pi/2_+}\ket{\pi/2_+}]_{2147}$. In {\it step}~(iv), qubit 6 is set to the state $\ket{+}$ after the polarizer in mode 6 and fused back into the resource with qubit 2. Next, in {\it step}~(v) qubit 4 is fused with qubit 3 from the state $[\ket{H}\ket{+}+\ket{V}\ket{-}]_{35}$. Three final fuse operations are required: qubit pairs (3,6),~(6,7) and (5,7) in {\it steps}~(vi),~(vii) and (viii) respectively. Qubit 7 is then measured-out in the computational basis with a polarizer. It is straightforward to show that when one photon is present in each output mode, leading to a {\it seven-fold} coincidence at detectors, the six-qubit graph shown in {\it step}~(ix) would have been generated~\cite{onewaye}. For performing one-way QC, the single-qubit measurements can be realized by replacing each of the detectors with a HWP-QWP-PBS chain and a detector at each of the output ports of the PBS~\cite{onewaye}. While state generation in this example is postselected,  
we note that a non-postselected approach~\cite{onewayt} would be needed for linking the Toffoli graphs up to larger optical resources for scalable QC. Thus, we stress that the steps given here are only an example. Efficient methods for fusing weighted graphs, based on adapting known heralded techniques~\cite{onewayt}, or additional photon degrees of freedom could help reduce the number of components and photons, while providing full scalability. Generating and controlling the graphs in systems such as ion traps~\cite{Stock} will require adaptation and an understanding of the dynamics of the system of interest.

We have introduced three compact graph states for performing a measurement-based Toffoli gate. Although no proof of optimality is given, in terms of qubit number, the techniques developed here should greatly aid the experimental demonstration of complex algorithms on small graph resources in various physical systems. They should reduce the effects of noise and imperfections due to their size, thus benefiting overall performance~\cite{onewayt}.   
 
We thank M. Paternostro and T. Yamamoto, and acknowledge support from EPSRC, QIPIRC and JSPS.


\end{document}